\documentclass[%
 reprint,
 amsmath,amssymb,
 aps, 
pra,
]{revtex4-2}

\usepackage[english]{babel}
\usepackage{graphicx}
\usepackage{dcolumn}
\usepackage{bm}
\usepackage{hyperref}
\usepackage{xcolor}
\usepackage{subcaption}
\usepackage{amsmath}
\usepackage{amssymb}
\usepackage{mathtools}
\usepackage{commath}
\usepackage{todonotes}
\usepackage{hhline}

\newcommand{\bhat}{\ensuremath{\boldsymbol{\hat{b}}}}
\newcommand*\bs[1]{\mathop{}\!\boldsymbol{#1}}

\def\email#1{Email address for correspondence: #1}

\begin{document}

\preprint{APS/123-QED}

\title{Saturation of magnetised plasma turbulence by propagating zonal flows}

\author{R. Nies$^{1,2}$}
\author{F. Parra$^{1,2}$}
\author{M. Barnes$^{3}$}
\author{N. Mandell$^{2}$}
\author{W. Dorland$^{4}$}
\thanks{Prof. W. Dorland passed away since submission of this article.}
\email{ richard.nies@physics.ox.ac.uk }
\affiliation{%
  $^{1}$Department of Astrophysical Sciences, Princeton University, Princeton, NJ 08543, USA\\
  $^{2}$Princeton Plasma Physics Laboratory, Princeton, NJ 08540, USA\\
  $^{3}$Rudolf Peierls Centre for Theoretical Physics, University of Oxford, Oxford OX1 3NP, United Kingdom\\
  $^{4}$Department of Physics, University of Maryland, College Park, MD 20742, USA
}%


\date{\today}

\begin{abstract}
Strongly driven ion-scale turbulence in tokamak plasmas is shown to be regulated by a new propagating zonal flow mode, the toroidal secondary mode, which is nonlinearly supported by the turbulence. The mode grows and propagates due to the combined effects of zonal flow shearing and advection by the magnetic drift. Above a threshold in the turbulence level, small-scale toroidal secondary modes become unstable and shear apart turbulent eddies, forcing the turbulence level to remain near the threshold. This threshold condition is used to derive scaling laws for the turbulent heat flux, fluctuation spectra, and zonal flow amplitude, which are validated in nonlinear gyrokinetic simulations and explain previous experimental observations.

\end{abstract}

\maketitle

\section{Introduction}
\label{sec:Intro}

Efforts to achieve the high temperatures required for thermonuclear fusion in magnetically confined plasmas are stymied by the large heat flux caused by turbulent mixing. The turbulent fluctuations are driven by micro-instabilities, most notably the ion temperature gradient (ITG) instability \cite{coppi_instabilities_1967}. The saturation level of such ion-gyroradius-scale modes crucially depends on zonal flows (ZFs) \cite{lin_turbulent_1998, dimits_comparisons_2000}, flow bands that are nonlinearly generated by the turbulence \cite{rogers_generation_2000} and shear apart turbulent eddies. In toroidal plasmas, the linear ZF physics allows for stationary ZFs \cite{rosenbluth_poloidal_1998, hinton_dynamics_1999} and rapidly oscillating geodesic acoustic modes (GAMs) \cite{winsor_geodesic_1968, conway_geodesic_2021}. 

In this article, we present the toroidal secondary mode (TSM), a ZF mode that propagates radially due to toroidal geometric effects and nonlinear interaction with the turbulence. In Section~\ref{sec:TSMs}, we show that the TSM explains the dynamics of small-scale ZFs in turbulence simulations, and we elucidate the physics of TSMs in Section~\ref{sec:TSM_physical_mechanism}. Furthermore, we demonstrate the TSM's relevance to turbulence saturation in Section~\ref{sec:energy_transfer} and show how the application of the critical balance conjecture to ITG turbulence \cite{barnes_critically_2011} must be altered to account for the TSM in Section~\ref{sec:GCB}. We derive scalings of the fluctuation amplitude and length scales, which are in agreement with turbulence simulations and explain experimental observations \cite{ghim_experimental_2013}. Finally, we conclude in Section~\ref{sec:Discussion}.

\section{Tokamak turbulence and gyrokinetics}

We begin with a brief introduction to the topic of tokamak turbulence and its description by gyrokinetic theory. To avoid fast parallel losses, the magnetic field $\bs{B}$ of tokamaks is made to lie on nested toroidal `flux surfaces'. The plasma rapidly reaches local thermodynamic equilibrium and is thus approximately Maxwellian, with the density $n_s$ and temperature $T_s$ of each species $s$ uniform on flux surfaces. The gradients between the hot dense core and the cold dilute edge drive turbulent fluctuations which cause transport across flux surfaces. These fluctuations evolve slowly compared to the Larmor gyration frequency $\Omega_s = e_s B /m_s$, with $m_s$ the particle mass and $e_s$ the charge, and $B$ the magnetic field strength. The fast gyratory motion may thus be averaged over and the turbulence is modelled using gyrokinetics \cite{catto_linearized_1978}, which describes rings of charge centred on the gyrocentre position $\bs{R}$, shifted from the particle position $\bs{r}$ by the gyroradius vector.

In the tokamak core, the fluctuations in the distribution function $f_s$ are small compared to the Maxwellian background $F_{\mathrm{M}s}$, i.e., $\delta f_s = f_s - F_{\mathrm{M}s} \ll F_{\mathrm{M}s}$, and they are highly anisotropic. Indeed, their typical length scales parallel to the magnetic field are of the order of the tokamak size, as measured e.g. by the major radius $R$, while in the perpendicular direction they are on the much smaller gyroradius scale $\rho_s = v_{Ts}/\Omega_s \ll R$, with $v_{Ts}=(2 T_s/m_s)^{1/2}$ the thermal speed. The fluctuations may thus be described locally in a flux-tube domain \cite{beerFieldalignedCoordinatesNonlinear1995}. In this article, we consider tokamak flux surfaces with circular cross sections and we choose the radial coordinate to be $x=r-r_0$, with $r$ the radial distance from the magnetic axis and $r_0$ the radius of the flux surface on which the flux tube is centred. As $x$ labels flux surfaces, its gradient is parallel to the density and temperature gradients. The binormal coordinate $y$ and the parallel coordinate $\theta$ determine the position within a flux surface: $y$, like $x$, is perpendicular to $\bs{B}$, whereas $\theta$ gives the location along $\bs{B}$. Within the thin flux tube, background quantities such as the magnetic field magnitude do not vary much in the direction perpendicular to the magnetic field. For this reason, it is convenient to Fourier analyse the directions $x $ and $y$. The wavenumbers $k_x$ and $k_y$ are the wavenumbers that correspond to $x$ and $y$.

Finally, we will use repeatedly the flux-surface average $\langle A \rangle_{y\theta}$ of a quantity $A$, given by its Jacobian-weighted average in $y$ and $\theta$. The Jacobian of the transformation to $(x,y,\theta)$ is proportional to $1/\bs{B}\cdot\nabla\theta$. We also define the zonal and nonzonal components of $A$ through $A^\mathrm{Z} \equiv \langle A \rangle_y = A - A^\mathrm{NZ}$.

Unless specified otherwise, the gyrokinetic simulations presented in this article are performed using \texttt{stella} \cite{barnes_stella_2019}. The simulations model a flux surface at half radius $r_0=a/2$ of the Cyclone Base Case (CBC) \cite{dimits_comparisons_2000}, a tokamak with inverse aspect ratio $R/a=2.8$, where $a$ is the minor radius. The safety factor $q$ gives the magnetic field-line pitch and sets the connection length $L_\parallel = q R$, with $2\pi L_\parallel$ the distance along the field line corresponding to one poloidal turn. The safety factor is varied between simulations, as is the ion temperature gradient $R/L_T\equiv -R \;\mathrm{d}\ln T_i/\mathrm{d}r$, while the density gradient and other quantities are held fixed at the usual CBC values. These are listed in Appendix~\ref{app:details_GK_sims}, alongside various numerical parameters.

We consider electrostatic and collisionless ion-gyroradius-scale fluctuations for a single ion species with charge $e_i = Z_i e$ and $e$ the proton charge. The gyrokinetic equation may be written in terms of the non-adiabatic part of the perturbed ion distribution function $h_i = \delta f_i + F_{\mathrm{M}i} e_i \varphi/T_i$, using the particle energy and magnetic moment as velocity-space coordinates,
\begin{eqnarray} \label{eq:gyrokinetic_eq_realspace_gs}
    & \partial_t \left( h_i-  F_{\mathrm{M}i} e_i \overline{\varphi} /T_i  \right) + \left(v_\parallel \bhat + \bs{\tilde v}_{M}  \right) \cdot \nabla h_i\nonumber\\
    & \; + \overline{\bs{v}_E} \cdot\nabla \left( F_{\mathrm{M}i} + h_i\right) = 0,
\end{eqnarray}
where the fluctuating electrostatic potential $\varphi$ is self-consistently determined by quasineutrality
\begin{eqnarray}\label{eq:quasineutrality}
    \frac{T_i}{e_i n_i}\int\mathrm{d}^3 v\; \overline{h_i}  = \tau \left( \varphi - \langle \varphi \rangle_{y\theta} \right) + \varphi,
\end{eqnarray}
with $\tau = T_i/Z_i T_e$ and $\bhat = \bs{B}/B$. The overlines indicate gyro-averages, with the averages $\overline{\varphi}$ and $\overline{h_i}$ taken at fixed $\bs{R}$ and $\bs{r}$, respectively. Even though it does not appear explicitly because of the use of the non-adiabatic piece of the distribution function, equation \eqref{eq:quasineutrality} includes the classical polarisation response. The electrons were assumed in \eqref{eq:quasineutrality} to respond adiabatically only to potential variations within flux surfaces $\varphi-\langle \varphi \rangle_{y\theta}$, as they cannot move radially due to their small gyroradii \citep{dorlandGyrofluidTurbulenceModels1993}. The gyrokinetic equation includes advection by the magnetic drift $\bs{\tilde v}_M$ and the gyro-averaged $\bs{E}\times\bs{B}$ drift $\overline{\bs{v}_E} = \bhat\times\nabla\overline{\varphi}/B$. The latter determines the time-and-volume-averaged heat flux $\langle Q \rangle_{txy\theta}$, with $Q = Q_\parallel + Q_\perp$ and
\begin{equation} \label{eq:heat_flux}
    Q_{\parallel, \perp} = \frac{1}{\langle\; \abs{\nabla x} \rangle_\theta}\int\mathrm{d}^3 v \; \frac{m_i v_{\parallel,\perp}^2}{2} \overline{h_i} \bs{v}_E \cdot \nabla x ,
\end{equation}
which is of primary interest to ascertain the transport caused by turbulent fluctuations.

\begin{figure}
     \centering
     \begin{subfigure}[t]{0.495\columnwidth}
         \centering
          \includegraphics[width=\textwidth, trim={0.9cm 0.9cm 0.5cm 0.9cm},clip]{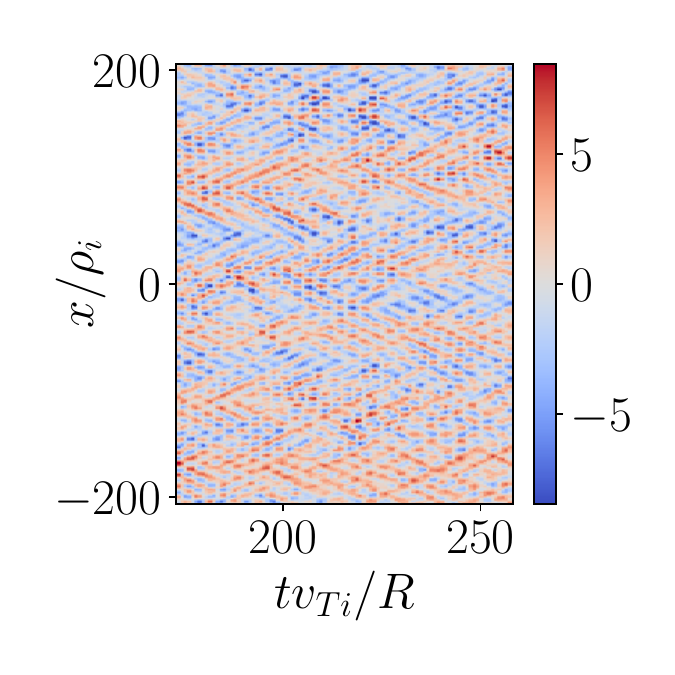}
     \end{subfigure}
     \begin{subfigure}[t]{0.485\columnwidth}
         \centering
          \includegraphics[width=\textwidth, trim={0.9cm 0.85cm 0.5cm 0.6cm},clip]{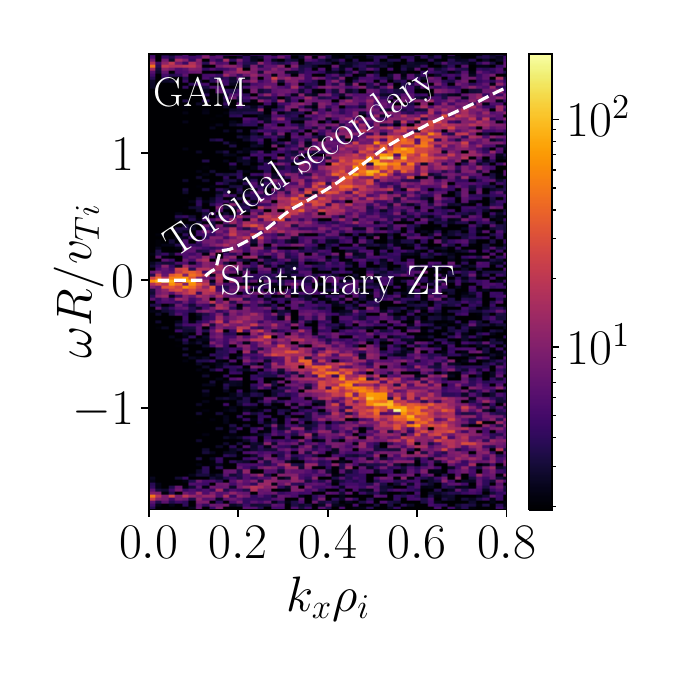}
     \end{subfigure}
    \caption{Normalised zonal flow velocity $\langle v_E^\mathrm{Z}/v_{Ti} \rangle_{\theta} R/\rho_i$ from saturated ITG turbulence far from marginality ($q=2.8, R/L_T = 10.4$) in real space (left) and the absolute value of its Fourier amplitude (right). Remarkably, the oscillation frequency of the secondary mode due to a primary streamer (dashed line) reproduces the frequency in the nonlinear turbulence simulation. The streamer has amplitude $v_{Ex}^P = \langle v_{Ex}^2\rangle_{y\theta}^{1/2} = \rho_i v_{Ti}/R$, is sinusoidal in the binormal direction with wavenumber $k_y^P \rho_i = 0.01$, is constant in $\theta$, and is bi-Maxwellian in velocity space with parallel temperature $T_\parallel^P = 0$ and a perpendicular temperature to density ratio of $T_\perp^P/T_i = 3 n^P/n_i$. We note that the details of the primary do not substantially affect the secondary mode frequency (see \cite{nies_theory_2025-1}).}
    \label{fig:ZF_real_Fourier}
\end{figure}

\section{Propagating zonal flows}
\label{sec:TSMs}

In gyrokinetic simulations of strongly driven ITG turbulence, ZF activity is apparent at multiple scales, see Fig.~\ref{fig:ZF_real_Fourier}, showing large-scale stationary ZFs and small-scale propagating ZFs. The corresponding Fourier spectrum in Fig.~\ref{fig:ZF_real_Fourier} exhibits stationary ZFs and GAMs at large scales $k_x \rho_i \sim 0.1$ and a new propagating ZF mode, the TSM. It is predominant at smaller scales $k_x \rho_i \sim 0.5$ and has a frequency $\omega \sim k_x v_{Mx}$ set by the radial magnetic drift velocity $v_{Mx} \equiv \rho_i v_{Ti}/R$, defined such that
\begin{equation}\label{eq:vMx}
    \tilde v_{Mx} = v_{Mx} \sin\theta \frac{v_\parallel^2 + v_\perp^2/2}{v_{Ti}^2}
\end{equation}
in a large aspect ratio tokamak of circular cross section.

\begin{figure}
    \centering
    \includegraphics[width=0.99\columnwidth, trim={0.3cm 0.9cm 1.1cm 0.9cm},clip]{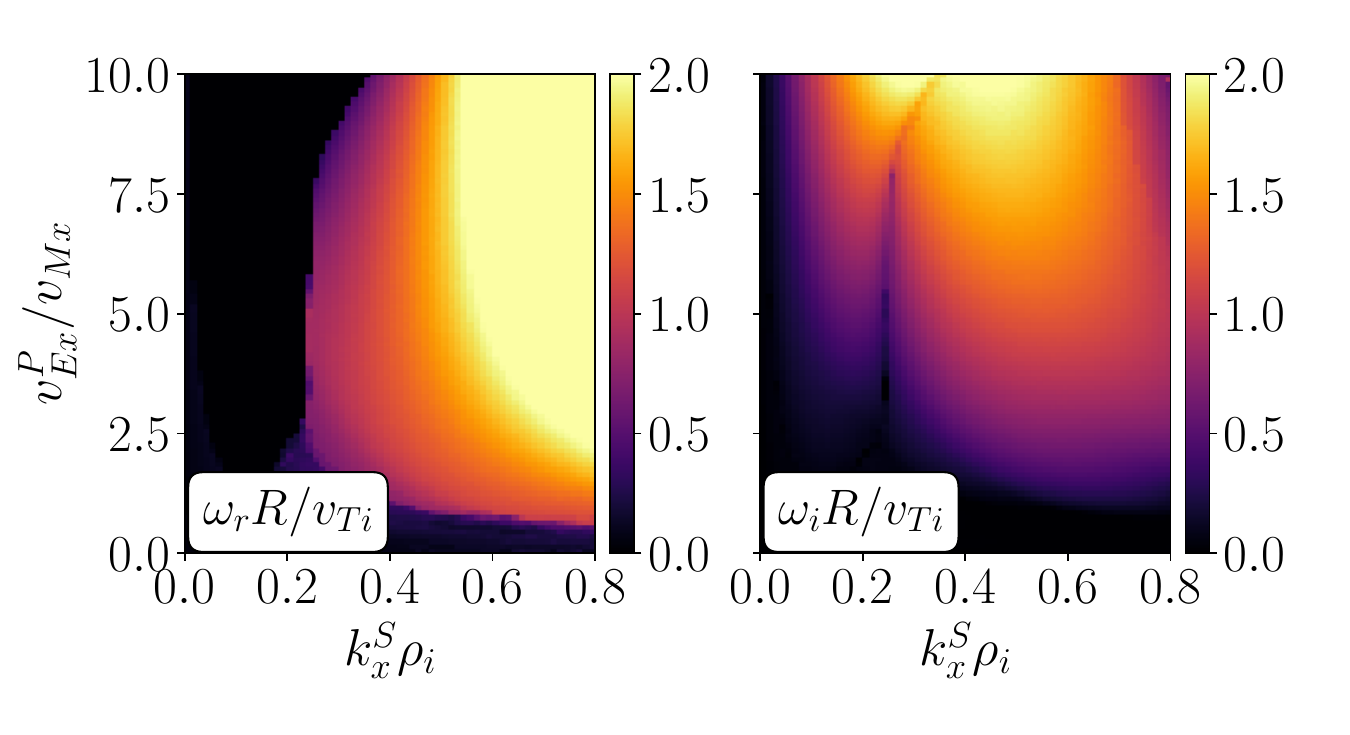}
	\caption{Oscillation frequency (left) and growth rate (right) of secondary modes with radial wavenumber $k_x^S$ due to a primary streamer mode with varying amplitude, measured by $v_{Ex}^P = \langle v_{Ex}^2\rangle_{y\theta}^{1/2}$. The primary is a linear ITG mode with binormal wavenumber $k_y \rho_i = 0.05$, at $q=1.4$ and $R/L_T=13.9$.}
    \label{fig:toroidal_secondary_GKsec_2D}
\end{figure}

Compared to the small-scale and fast-oscillating TSM, the largest turbulent eddies have long radial wavelengths and evolve slowly in time, as will be shown below. The TSM may therefore be studied by considering a secondary mode growing and propagating over a background stationary ($\partial_t=0$) streamer ($\partial_x=0$) mode, referred to as the primary mode and taken to be representative of the turbulence. This secondary model was first considered by \cite{rogers_generation_2000}, who ignored the magnetic drift and parallel streaming in \eqref{eq:gyrokinetic_eq_realspace_gs} and derived a purely growing mode ($\omega_r = 0$). As shown in Fig.~\ref{fig:toroidal_secondary_GKsec_2D}, gyrokinetic simulations of the secondary instabilities of a streamer exhibit small-scale oscillating ($\omega_r \neq 0$) ZFs; a purely growing mode is observed only at large primary drive or long ZF wavelengths. In \cite{nies_theory_2025-1}, the dispersion relation for TSMs is derived and validated with gyrokinetic simulations -- in this article, we instead focus on the phenomenology and physical mechanism of the TSM, as well as its effect on the turbulence.

\section{Physical mechanism causing the toroidal secondary mode}
\label{sec:TSM_physical_mechanism}

To understand how the TSM drives ZFs and propagates, we consider the vorticity equation at long perpendicular wavelengths,
\begin{eqnarray} \label{eq:vorticity}
    & n_i \partial_t \left\langle \rho_i \abs{\nabla x} \partial_x v_E^\mathrm{Z} \right\rangle_{\theta} = \\
    &v_{Ti} \partial_x \left\langle \int\mathrm{d}^3 v\;  \left(  \frac{v_\perp^2}{2\Omega_i^2} \nabla x \cdot \nabla \bs{v}_E \cdot \nabla h_i + \tilde v_{Mx}h_i \right)\right\rangle_{y\theta} \nonumber,
\end{eqnarray}
which is derived by taking the time derivative and flux-surface average of \eqref{eq:quasineutrality}, see Appendix~\ref{app:derivation_vorticity}. The ZF velocity $v_E^\mathrm{Z}=\abs{\nabla x}\partial_x \langle \varphi\rangle_y/B$ (or relatedly the zonal vorticity $\langle (\nabla \times \bs{v}_E) \cdot \bhat \rangle_y \approx \abs{\nabla x} \partial_x v_E^\mathrm{Z} $) evolves due to nonlinear stresses and due to the $\tilde{v}_{Mx}$-induced Stringer-Winsor (SW) force \cite{stringer_diffusion_1969, winsor_geodesic_1968, hassam_spontaneous_1993, hallatschek_transport_2001}. Because the magnetic drift points radially outward above the tokamak midplane and inward below it, as reflected in the $\sin\theta$ of \eqref{eq:vMx}, the SW force requires up-down asymmetry in $(P+e_i n_i \varphi)^\mathrm{Z}$, where
\begin{equation}
    P \equiv \int\mathrm{d}^3 v\; \frac{m_i}{2}\left( v_\parallel^2 + \frac{v_\perp^2}{2} \right) \delta f_i
\end{equation}
is a fluctuating pressure-like quantity.

The nonlinear stresses in \eqref{eq:vorticity} have been the focus of most studies of ZF growth, including the secondary mode theory of \cite{rogers_generation_2000}, subsequent generalisations by \citep{plunkGyrokineticSecondaryInstability2007}, \citep{chen_excitation_2000}, and \citep{plunk_nonlinear_2017}, and quasilinear models \citep[e.g.][]{diamond_dynamics_2001, parkerZonalFlowPattern2013, zhuWavekineticApproachZonalflow2021}. The magnetic drift was neglected in these studies, precluding SW force effects on the ZFs. 

\begin{figure}
     \centering
          \includegraphics[width=\columnwidth, trim={1.8cm 0.1cm 1.9cm 1.1cm},clip]{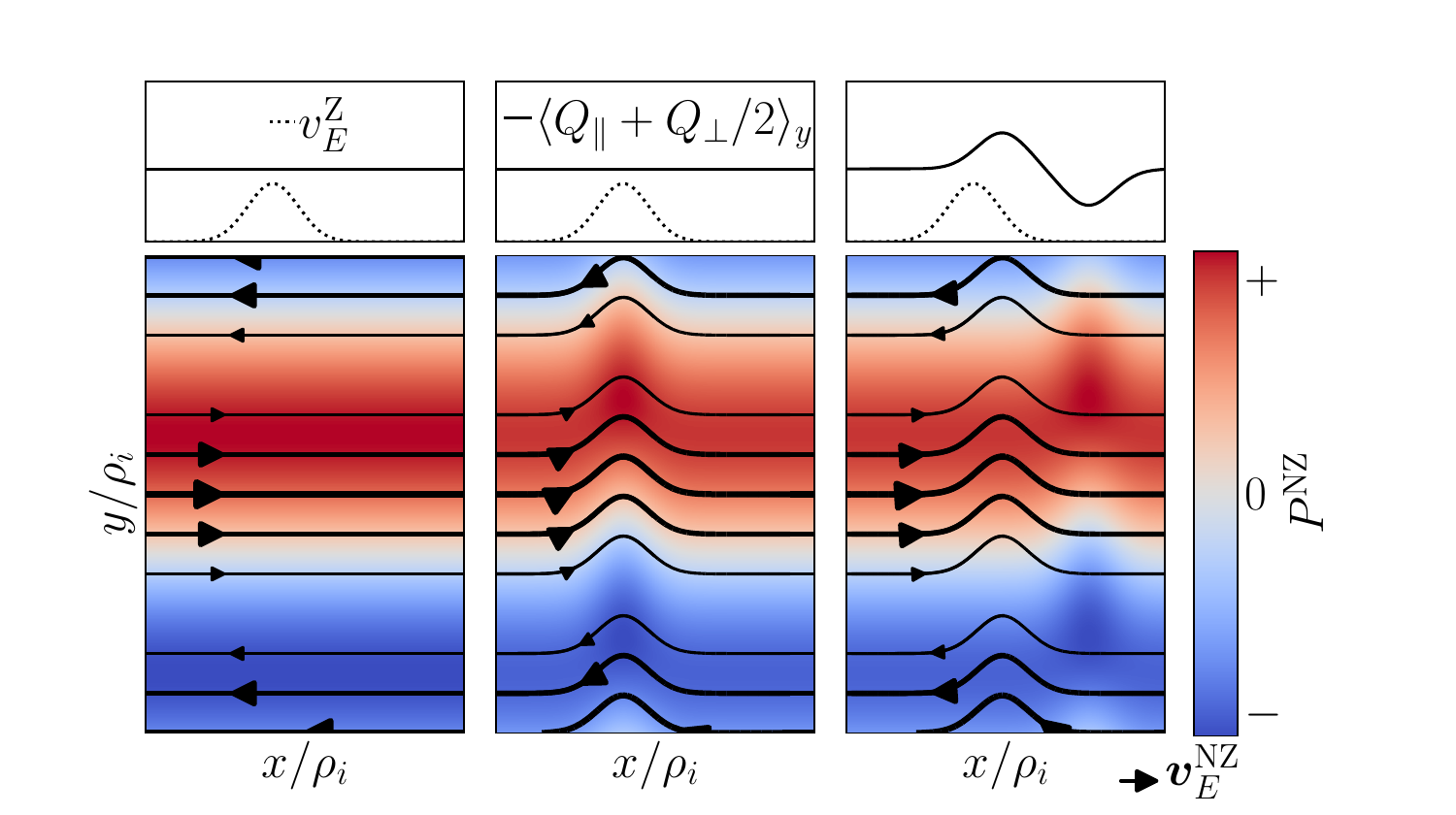}
		 \caption{Generation of $(P+e_i n_i \varphi)^\mathrm{Z}$ asymmetry by nonlinear heat flux, illustrated above the tokamak midplane where $\tilde v_{Mx}>0$: starting from a streamer mode (left), the $v_E^\mathrm{Z}$ perturbation shears the $\bs{v}_E^\mathrm{NZ}$ and $P^\mathrm{NZ}$ contours equally, leaving the heat flux $\langle Q_\parallel + Q_\perp/2 \rangle_y \propto \langle v_{Ex}^\mathrm{NZ} P^\mathrm{NZ} \rangle_y$ \eqref{eq:heat_flux} unchanged (middle). Due to the advection by the velocity-dependent $\tilde v_{Mx}$ \eqref{eq:vMx}, a relative displacement (phase shift) between $\bs{v}_E^\mathrm{NZ}$ and $P^\mathrm{NZ}$ ensues, leading to a radial modulation of $\langle Q_\parallel+Q_\perp/2 \rangle_y$ (right). This mechanism is reversed below the midplane where $\tilde v_{Mx}<0$, causing an up-down asymmetry in the heat flux and thence in $(P+e_i n_i \varphi)^\mathrm{Z}$.}
    \label{fig:TSM_physics}
\end{figure}

In the case of the TSM, nonlinear stresses are subdominant to the SW force in \eqref{eq:vorticity}. In a torus, asymmetry in $(P+e_i n_i \varphi)^\mathrm{Z}$ is generated linearly from a ZF due to geodesic curvature. Balancing the resulting SW force with the ZF inertia in \eqref{eq:vorticity} (the term with $\partial_t$) leads to GAMs. For the TSM, the linearly induced SW force is instead balanced with a nonlinearly generated $(P+e_i n_i \varphi)^\mathrm{Z}$ asymmetry, illustrated in Fig.~\ref{fig:TSM_physics}. The ZF shearing and advection by the velocity-dependent $\tilde v_{Mx}$ together cause a phase shift between nonzonal pressure and potential fluctuations, such that $\partial_x \langle Q_\parallel+Q_\perp/2 \rangle_y \neq 0$. The resulting up-down asymmetric heat flux generates $(P+e_i n_i \varphi)^\mathrm{Z}$ asymmetry on flux surfaces neigbouring the ZF perturbation, causing the mode to propagate.

This nonlinear generation of $(P+e_i n_i \varphi)^\mathrm{Z}$ asymmetry, which must generally be modelled accounting for both toroidal geometric and kinetic effects \cite{nies_theory_2025-1}, enables ZF growth by a mechanism entirely different to that from nonlinear stresses. We note that the TSM mechanism is also distinct from that of \cite{hallatschek_transport_2001, itoh_excitation_2005}, who consider pressure asymmetry generation due to a combination of ZF shearing and magnetic shear.

The physical picture of the TSM outlined above neglected parallel streaming, which acts to short-circuit up-down asymmetries in $(P+e_i n_i \varphi)^\mathrm{Z}$. Therefore, the TSM is only found at short radial wavelengths, where $\omega \sim k_x v_{Mx} \gtrsim v_\parallel \bhat \cdot \nabla \sim v_{Ti}/qR$. At longer wavelengths, the ZFs do not oscillate, as seen in the gyrokinetic simulations of the secondary (Fig.~\ref{fig:toroidal_secondary_GKsec_2D}) or of the turbulence (Fig.~\ref{fig:ZF_real_Fourier}). Furthermore, the TSM requires a sufficiently large primary drive $v_{Ex}^P \gtrsim v_{Mx}$ to become unstable, as shown in Fig.~\ref{fig:toroidal_secondary_GKsec_2D}. Finally, the TSM is stabilised by finite Larmor radius effects, such that the growth rate peaks at $k_x \rho_i \sim 0.5$ in Fig.~\ref{fig:toroidal_secondary_GKsec_2D}, explaining the prominent ZFs at this scale in turbulence simulations (Fig.~\ref{fig:ZF_real_Fourier}).

\section{Zonal flows and turbulence saturation}
\label{sec:energy_transfer}

\begin{figure}
    \centering
    \includegraphics[width=\columnwidth, trim={0.9cm 0.65cm 0.9cm 0.9cm},clip]{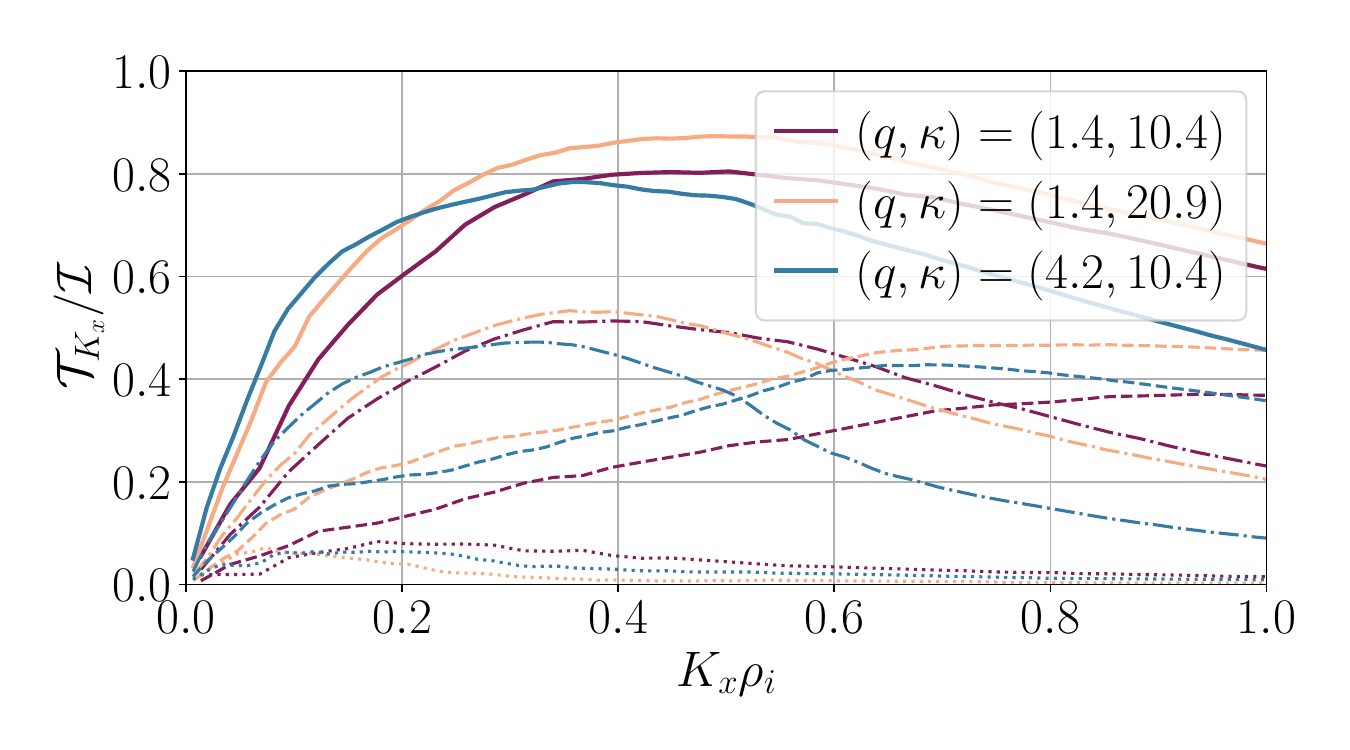}
	\caption{Energy transfer rate $\mathcal{T}_{K_x}$ \eqref{eq:energyflux} normalised by the total injection rate $\mathcal{I}=\lim_{K_x\rightarrow \infty} \mathcal{I}_{K_x}=\langle Q \rangle_{xy\theta}/L_T$ in turbulence simulations. The contributions to the transfer rate (solid) are obtained by decomposing $\bs{v}_E$ in \eqref{eq:energyflux} into its nonzonal (dashed), small-scale $\abs{k_x \rho_i} > 0.3$ zonal (dash-dotted), and large-scale $\abs{k_x \rho_i} \leq 0.3$ zonal (dotted) components. Different colours correspond to different simulations.}
    \label{fig:energyflux}
\end{figure}

To ascertain the role of the various ZF modes in setting the turbulence saturation level, we consider the energy transfer in $k_x$ caused by the nonzonal $\bs{E}\times\bs{B}$ flow and by ZFs at different scales in Fig.~\ref{fig:energyflux}. We first define the low-pass-filtered distribution
\begin{eqnarray}
    h_{i, K_x} \equiv \int_{-K_x}^{K_x}\mathrm{d}k_x\; \hat h_i(k_x) e^{i k_x X},
\end{eqnarray}
where the Fourier modes $\hat h_i(k_x)$ are normalised such that $h_i = \lim_{K_x \rightarrow \infty} h_{i, K_x}$. Then, the low-pass filtered gyrokinetic free energy
\begin{equation}
    \mathcal{E}_{K_x} \equiv \sum_s T_s \left\langle \int\mathrm{d}^3 v\; \frac{\left( \delta f_{s,K_x} \right)^2}{2 F_{\mathrm{M}s}} \right\rangle_{xy\theta}
\end{equation}
may be shown to evolve as 
\begin{equation}
    \partial_t \mathcal{E}_{K_x} + \mathcal{T}_{K_x} = \mathcal{I}_{K_x} - \mathcal{D}_{K_x},  
\end{equation}
with the energy transfer rate (note the gyro-average is over $h_i$ and $h_{i, K_x}$)
\begin{eqnarray}\label{eq:energyflux}
    \mathcal{T}_{K_x} = \left\langle T_i \int\mathrm{d}^3 v\; \overline{ \frac{h_{i, K_x}}{F_{\mathrm{M}i}} \bs{v}_E \cdot \nabla  h_i} \right\rangle_{xy\theta},
\end{eqnarray}
the injection rate
\begin{equation}
    \mathcal{I}_{K_x} = L_T^{-1}  \left\langle \int\mathrm{d}^3 v \; \frac{m_i v^2}{2} \overline{h_{i, K_x}} \bs{v}_E \cdot \nabla x \right\rangle_{xy\theta},
\end{equation}
and a dissipation rate $\mathcal{D}_{K_x}$ due to collisions or numerical dissipation.

In Fig.~\ref{fig:energyflux}, the small-scale ZFs are shown to transfer most of the free energy from large to small radial scales, as they are the largest contributors to $\mathcal{T}_{K_x}$ at wavelengths below the TSM scale $K_x \rho_i \sim 0.5 - 0.7$; for larger $K_x$ the nonzonal contribution is dominant. The contribution to $\mathcal{T}_{K_x}$ from the large-scale ZFs is subdominant \footnote{We note that numerical experiments artificially removing the ZFs at small $k_x$ were found to alter the turbulence saturation level. The interpretation of such numerical experiments is ultimately difficult -- after all, removing the long-wavelength ZFs while keeping the short-wavelength ones requires forces on the plasma with very particular properties.}. We note that the maximum value of $\mathcal{T}_{K_x}/\mathcal{I}$ is close to unity at $K_x \rho_i \approx 0.5$; therefore, most of the energy is injected at large scales $K_x \rho_i < 0.5$ and dissipated at small scales $K_x \rho_i > 0.5$.

\section{Turbulence scaling laws}
\label{sec:GCB}

Strongly driven ITG turbulence in tokamaks was previously modelled \cite{barnes_critically_2011} using the critical balance conjecture \cite{goldreich_toward_1995}, which posits the nonlinear transfer rate $\omega_\mathrm{NL}$ and parallel propagation rate $\omega_\parallel$ of the saturated turbulence to be of the same order: at every scale, $\omega_\mathrm{NL} \sim v_{Ex}/l_x \sim \omega_\parallel \sim v_{Ti} / l_\parallel $, with the rates estimated from the gyrokinetic equation \eqref{eq:gyrokinetic_eq_realspace_gs}. Here, $l_x, l_y, l_\parallel$ are the radial, binormal, and parallel length scales of the turbulence. The turbulence outer scale, denoted by `o' superscripts, is defined (analogously to the stirring scale in fluid turbulence) to be the scale at which the nonlinear transfer rate balances the energy injection rate $\omega_\star^T$, i.e. $\omega_\mathrm{NL}^o  \sim \omega_\star^{T,o} \sim (v_{Ti}/L_T) \rho_i/l_y^o$. Here, we have followed \cite{barnes_critically_2011} in assuming $e_i \varphi^o / T_i \sim h_i/F_{Mi}$, which by quasineutrality \eqref{eq:quasineutrality} is equivalent to assuming different velocity moments of $h_i$ scale in the same way (this is generally not satisfied for linear modes). This assumption may be justified by the inclusion of parallel streaming and magnetic drift advection rates in the outer scale balance (see below), as these terms couple velocity moments of $h_i$. Finally, we also follow \cite{barnes_critically_2011} in assuming the parallel outer length scale to be the connection length $l_\parallel^o \sim q R$.

\subsection{Marginality of toroidal secondary modes}

In \cite{barnes_critically_2011}, the turbulent eddies were assumed to be isotropic in the plane perpendicular to the magnetic field ($l_x \sim l_y$). We have found this assumption to be incorrect because the turbulence at the outer scale is anisotropic in this plane. We conjecture that the turbulence level is regulated to be near the marginal stability threshold of the TSM, $v_{Ex}\sim v_{Mx}$ at the outer scale (see Fig.~\ref{fig:toroidal_secondary_GKsec_2D}). 

\begin{figure}
     \centering
     \begin{subfigure}[t]{\columnwidth}
         \centering
          \includegraphics[width=\textwidth, trim={1.5cm 1.2cm 1.7cm 0.8cm},clip]{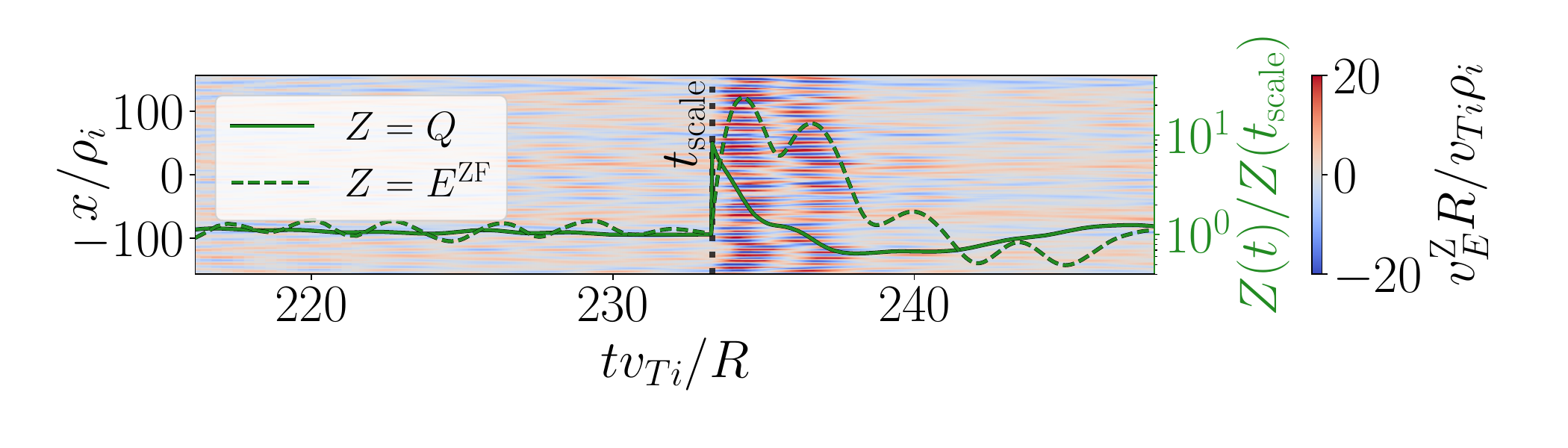}
		 \caption{At $t=t_\mathrm{scale}$, the nonzonal distribution is multiplied by $3$.}
         \label{fig:GCB_numerical_experiments_NZ_up}
     \end{subfigure}
     \begin{subfigure}[t]{\columnwidth}
         \centering
          \includegraphics[width=\textwidth, trim={1.5cm 1.2cm 1.7cm 0.4cm},clip]{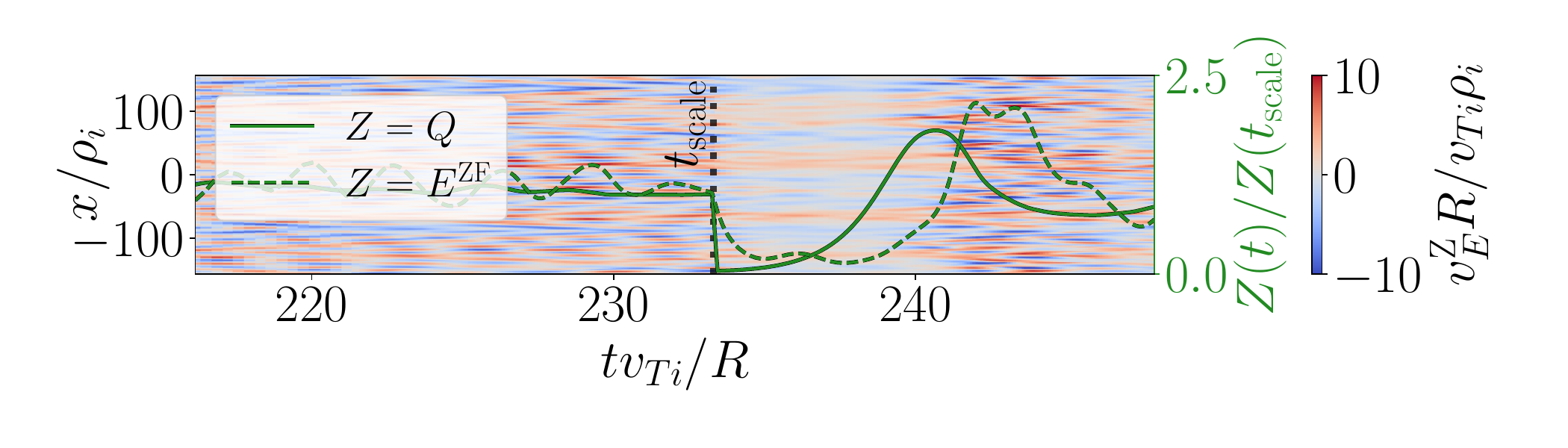}
		 \caption{At $t=t_\mathrm{scale}$, the nonzonal distribution is divided by $5$.}
         \label{fig:GCB_numerical_experiments_NZ_down}
     \end{subfigure}
    \caption{Numerical experiments rescaling the nonzonal distribution function. The rescaling occurs at $t=t_\mathrm{scale}$ (vertical dotted line), during the saturated phase of the \texttt{stella} turbulence simulation with $R/L_T=13.9$ and $q=1.4$. The plots show the ZF velocity (color coded) as a function of $x$ and $t$, and the temporal evolution of the nonlinear heat flux $Q$ (green solid) and the ZF energy $E^\mathrm{ZF}$ (green dashed) \eqref{eq:E_ZF}. The values of $Q$ and $E^\mathrm{ZF}$ are given on the right vertical axis.}
    \label{fig:GCB_numerical_experiments}
\end{figure}

This conjecture is supported by numerical experiments where the nonzonal components of the ion distribution function are artificially rescaled during the saturated phase of a turbulence simulation, see Fig.~\ref{fig:GCB_numerical_experiments}. By considering the time evolution as the system returns to the saturated state, we gain insights into the interplay between turbulence saturation and ZF dynamics.

We first show in Fig.~\ref{fig:GCB_numerical_experiments_NZ_up} that a sudden increase in the nonzonal fluctuation level leads to a rapid growth of small-scale ZFs, i.e. of TSMs. As the ZFs grow to a large amplitude, the nonzonal fluctuation level rapidly returns to its original value, as shown by the reduction in the nonlinear heat flux $Q$. The resulting decrease in the ZF drive then also causes the ZF energy $E^\mathrm{ZF}$ to return to its saturated value. Second, we show in Fig.~\ref{fig:GCB_numerical_experiments_NZ_down} that a sudden reduction in the nonzonal fluctuation level is followed by a rapid decrease in the ZF energy to a small (but finite \citep{rosenbluth_poloidal_1998, hinton_dynamics_1999}) level as the TSM is stabilised. Eventually, the turbulence level exceeds the saturated level, causing the TSMs to become unstable and thereby inducing a rapid growth in the ZF energy. The large ZF shearing then quenches the turbulence, which ultimately returns to its original saturation amplitude.

\subsection{Implications for critical balance}

Combining the TSM marginality condition $v_{Ex} \sim v_{Mx}$ with the critical balance conjecture at the outer scale, we obtain
\begin{equation}
    \omega_\mathrm{NL}^o \sim \omega_\parallel^o \sim \omega_\star^{T,o} \sim \omega_{Mx}^o,
\end{equation}
with $\omega_{Mx} \sim v_{Mx}/l_x$. This balance was previously observed experimentally \cite{ghim_experimental_2013} and can now be explained based on the TSM physics. We emphasise that we consider the balance at the outer scale only; we leave for future work the prediction of the turbulence spectra at smaller scales.

\begin{figure}
     \centering
     \begin{subfigure}[t]{0.49\columnwidth}
         \centering
          \includegraphics[width=\textwidth, trim={2.5 0.8cm 0.5cm 0.6cm},clip]{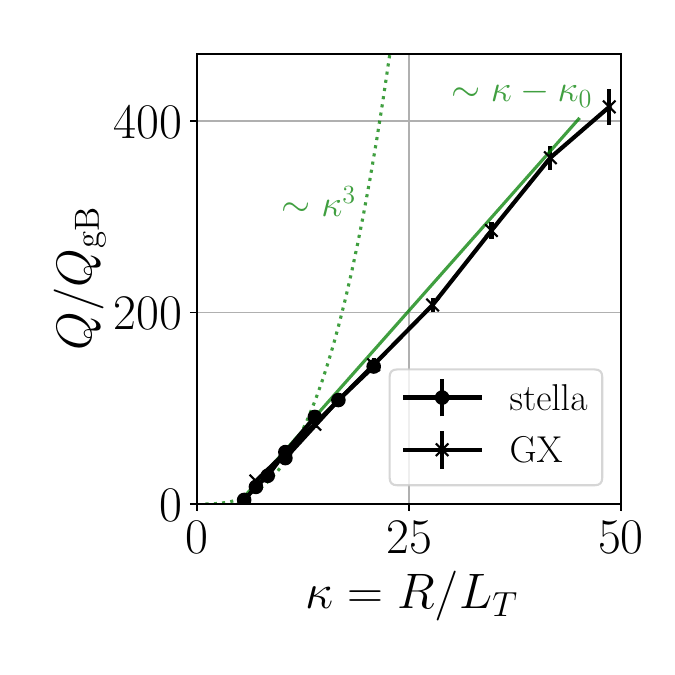}
		 \caption{Normalised heat flux}
         \label{fig:GCB_Q}
     \end{subfigure}
     \begin{subfigure}[t]{0.49\columnwidth}
         \centering
          \includegraphics[width=\textwidth, trim={2.5 0.8cm 0.5cm 0.6cm},clip]{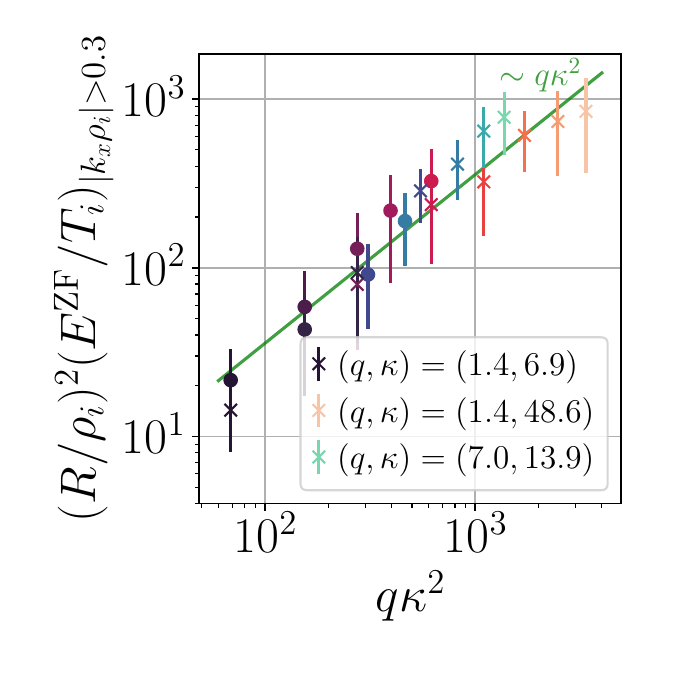}
		 \caption{Energy in small-scale ZFs}
         \label{fig:GCB_ZF}
     \end{subfigure}
     \begin{subfigure}[t]{0.49\columnwidth}
         \centering
          \includegraphics[width=\textwidth, trim={2.5 0.8cm 0.5cm 0cm},clip]{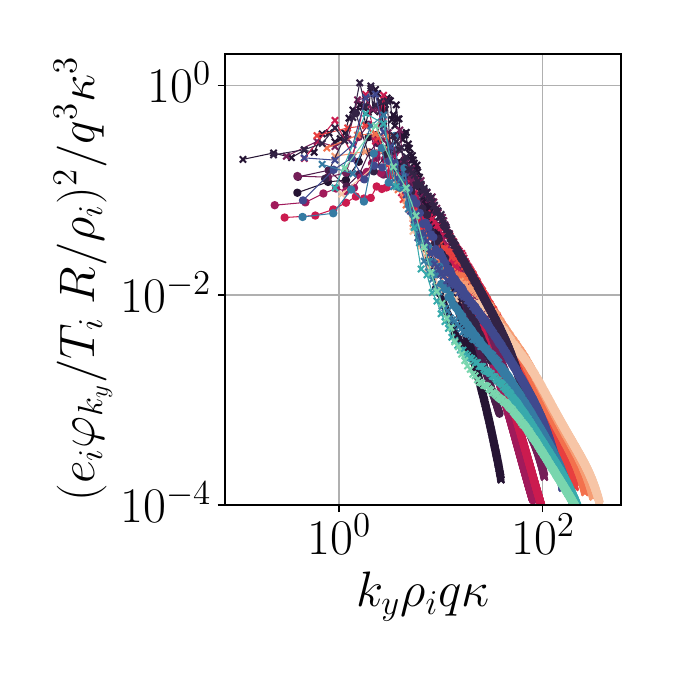}
		 \caption{Binormal spectrum}
         \label{fig:GCB_spectrum_ky}
     \end{subfigure}
     \begin{subfigure}[t]{0.49\columnwidth}
         \centering
          \includegraphics[width=\textwidth, trim={2.5 0.8cm 0.5cm 0cm},clip]{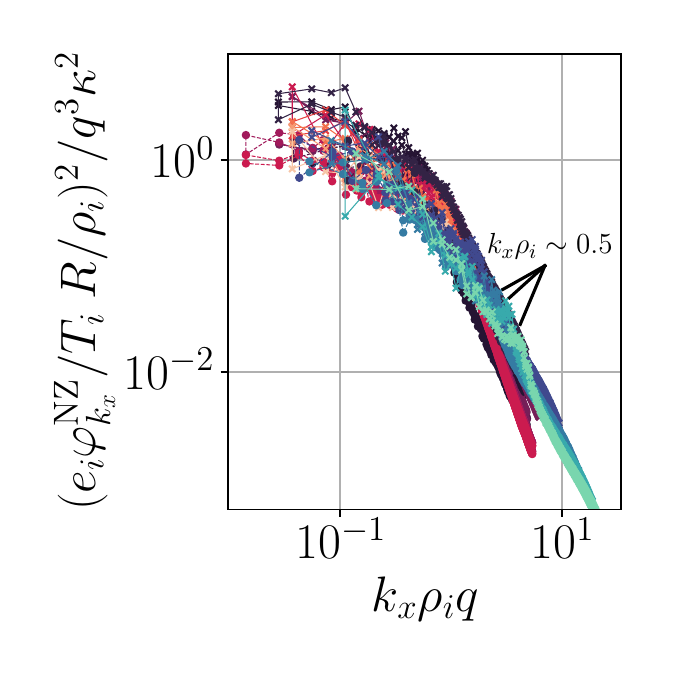}
          \caption{Radial spectrum}
         \label{fig:GCB_spectrum_kx}
     \end{subfigure}
    \caption{Validation of scalings \eqref{eq:GCB_scalings} in nonlinear gyrokinetic simulations using the codes \texttt{stella} \cite{barnes_critically_2011} (circles) and \texttt{GX} \cite{mandell_laguerrehermite_2018, mandellGXGPUnativeGyrokinetic2024} (crosses). The scaling of the heat flux with the background temperature gradient $\kappa=R/L_T$ \eqref{eq:GCB_scalings} is verified in (a) and the scaling of the zonal flow energy \eqref{eq:ZF_scaling} with $q$ and $\kappa$ is substantiated in (b). The nonzonal potential fluctuation spectra in (c-d) are rescaled according to \eqref{eq:GCB_scalings}. In (b-d), the cold colours (black$\rightarrow$blue$\rightarrow$cyan) and warm colours (black$\rightarrow$red$\rightarrow$yellow) correspond to the $q$ and $\kappa$ scans, respectively. The error bars in (a-b) correspond to the standard deviation in time. In (d), the peaks at the TSM scale $k_x \rho_i \sim 0.5$ are highlighted.}
    \label{fig:GCB}
\end{figure}

We obtain the following scalings for the turbulence perpendicular length scales, amplitude, and heat flux:
\begin{eqnarray}\label{eq:GCB_scalings}
    \frac{l_y^o}{\rho_i} \sim \frac{q R}{L_T},\; \frac{l_x^o}{\rho_i} \sim q, \; \frac{e_i \varphi^o}{T_i}\frac{R}{\rho_i} \sim \frac{q R}{L_T}, \; \frac{ Q }{Q_\mathrm{gB}}\sim  \frac{q R}{L_T},
\end{eqnarray}
with $Q_\mathrm{gB} = n_i T_i v_{Ti} (\rho_i/R)^2$ the gyro-Bohm heat flux. Notably, the heat flux is predicted to scale linearly with $R/L_{Ti}$, instead of the cubic scaling predicted assuming perpendicular isotropy \cite{barnes_critically_2011}\footnote{The study by \cite{barnes_critically_2011} also compared their scalings with gyrokinetic simulations of the CBC. However, due to computational limitations, the range of temperature gradients was limited to $R/L_T \leq 17.5$, where a cubic fit is a good approximation to an offset linear dependence, see Fig.~\ref{fig:GCB_Q}.}. This revised heat flux scaling is satisfied in simulations, as shown in Fig.~\ref{fig:GCB_Q}. We also verify the scalings \eqref{eq:GCB_scalings} for the turbulence amplitude and outer scale by considering the fluctuation spectra, see Figs.~\ref{fig:GCB_spectrum_ky}~and~\ref{fig:GCB_spectrum_kx} \footnote{We note that, although there is some variation between the curves, we do not observe a systematic trend in the deviation between the curves as the parameters $q$ and $\kappa$ are varied.}.

\begin{figure*}[!ht]
     \centering
     \begin{subfigure}[t]{0.3\textwidth}
         \centering
          \includegraphics[width=\textwidth, trim={2.5 0.8cm 0.5cm 1.4cm},clip]{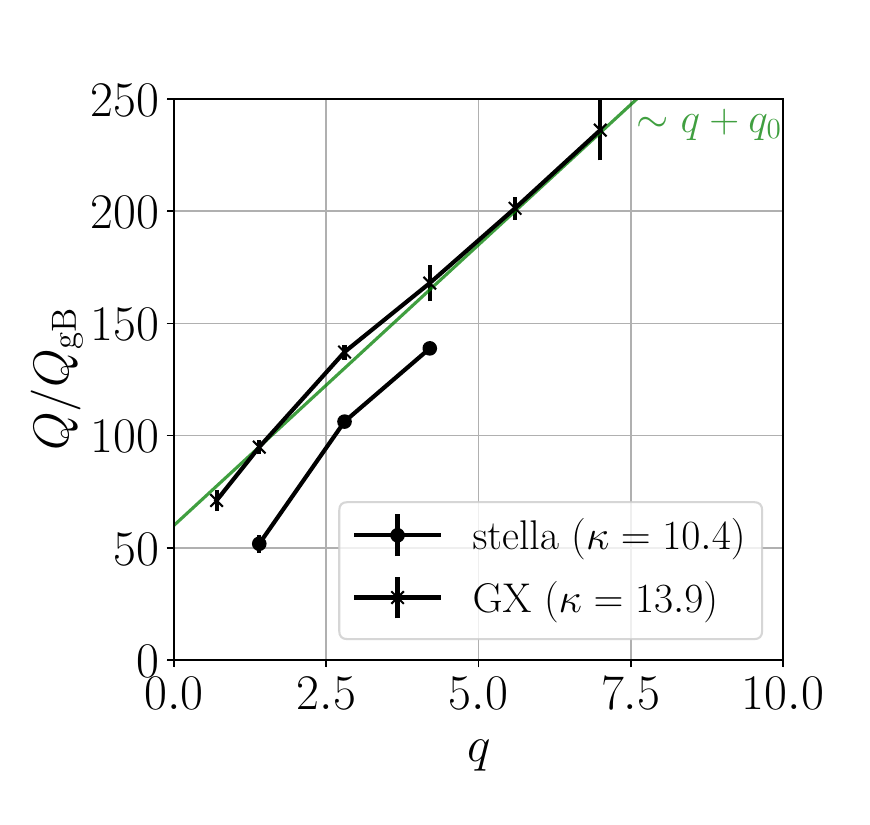}
		 \caption{Normalised heat flux}
         \label{fig:GCB_Q_q}
     \end{subfigure}
     \begin{subfigure}[t]{0.3\textwidth}
         \centering
          \includegraphics[width=\textwidth, trim={2.5 0.8cm 0.5cm 1.45cm},clip]{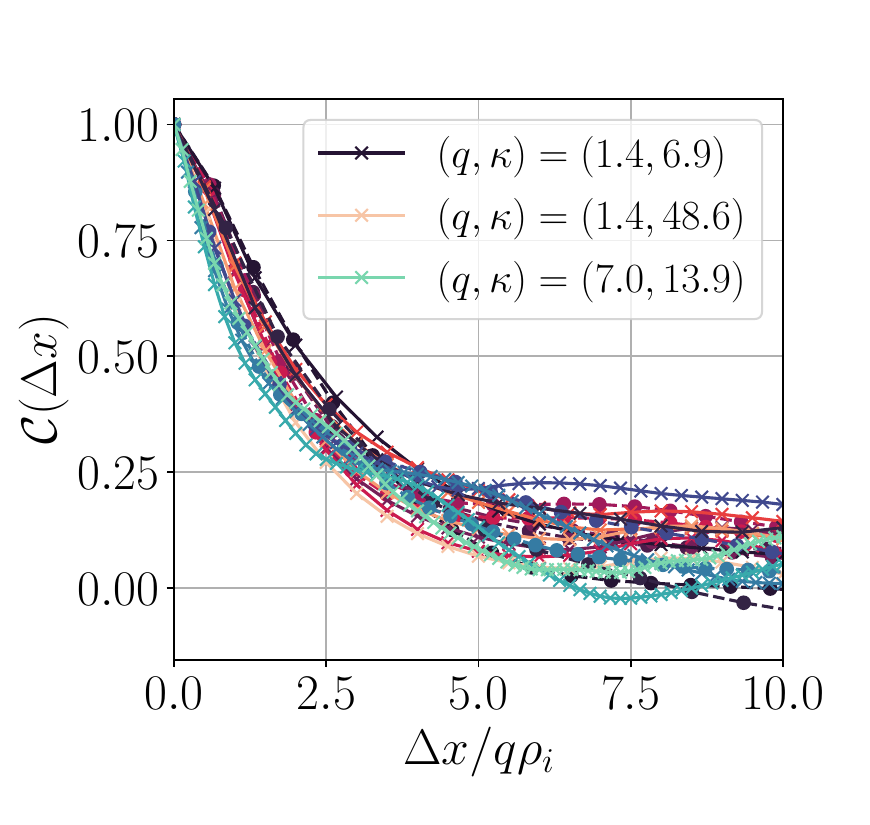}
		 \caption{Radial correlation function}
         \label{fig:GCB_correlation_x}
     \end{subfigure}
     \begin{subfigure}[t]{0.3\textwidth}
         \centering
          \includegraphics[width=\textwidth, trim={2.5 0.8cm 0.5cm 1.45cm},clip]{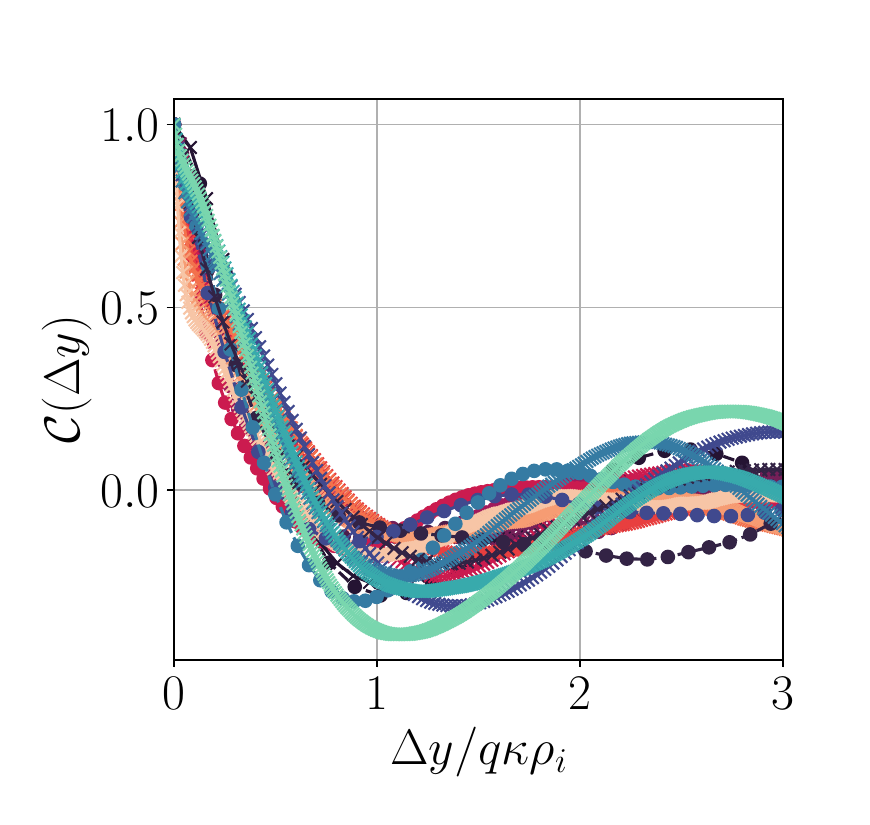}
		 \caption{Binormal correlation function}
         \label{fig:GCB_correlation_y}
     \end{subfigure}
     
     \begin{subfigure}[t]{0.4\textwidth}
         \centering
          \includegraphics[width=\textwidth, trim={2.5 0.8cm 0.5cm 0cm},clip]{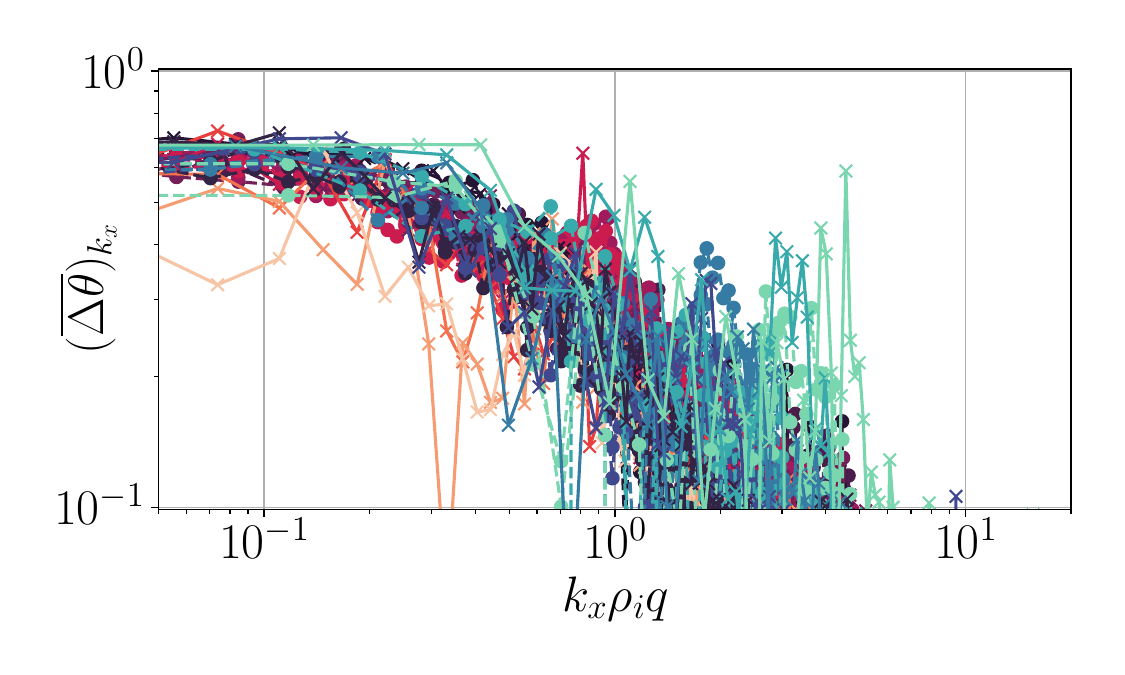}
		 \caption{Parallel correlation length as a function of $k_x$}
         \label{fig:GCB_correlation_kx}
     \end{subfigure}
     \hspace{0.7cm}
     \begin{subfigure}[t]{0.4\textwidth}
         \centering
          \includegraphics[width=\textwidth, trim={2.5 0.8cm 0.5cm 0cm},clip]{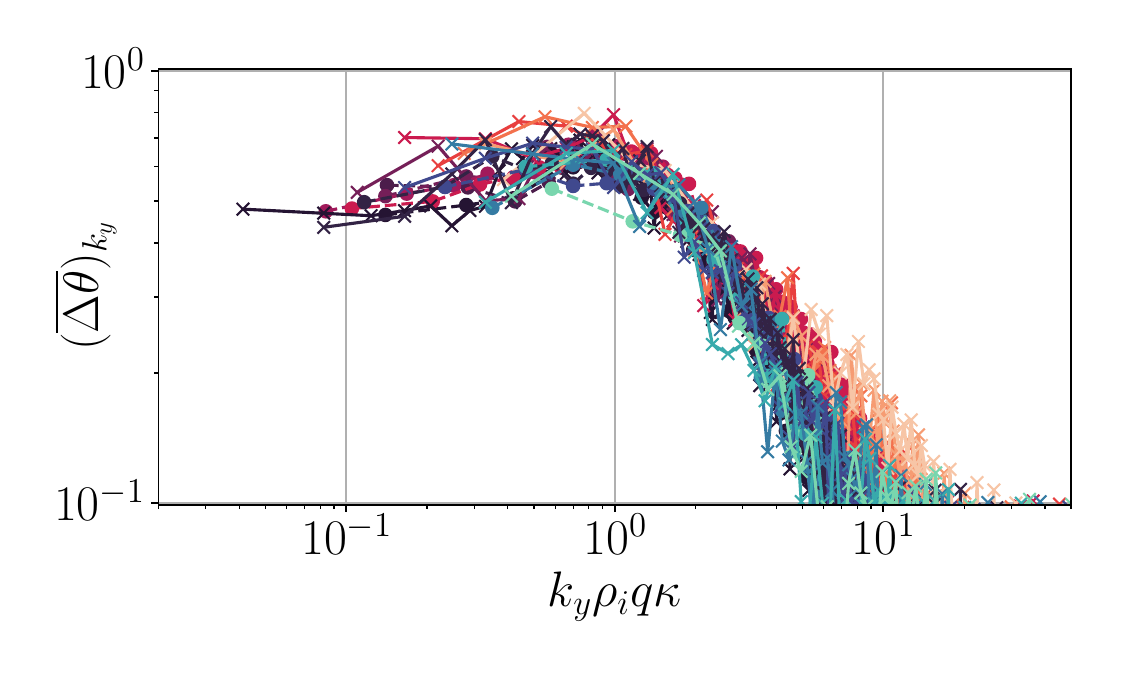}
          \caption{Parallel correlation length as a function of $k_y$}
         \label{fig:GCB_correlation_ky}
     \end{subfigure}
     \caption{Further validation of scalings \eqref{eq:GCB_scalings} in nonlinear gyrokinetic simulations using the codes \texttt{stella} \cite{barnes_critically_2011} (circles) and \texttt{GX} \cite{mandell_laguerrehermite_2018, mandellGXGPUnativeGyrokinetic2024} (crosses). The scaling of the heat flux $Q$ with the safety factor $q$ is verified in (a), the perpendicular correlation functions are shown in (b) and (c), and the parallel correlation lengths are shown in (d) and (e). The radial and binormal coordinates for the correlation functions (b,c) and lengths (d,e) are rescaled according to \eqref{eq:GCB_scalings}. The colour scheme of (b-e) is explained in the caption of Fig.~\ref{fig:GCB}.}
    \label{fig:GCB_extra}
\end{figure*}

The offset linear scaling of the heat flux $Q$ as a function of the safety factor $q$ is shown in Fig.~\ref{fig:GCB_Q_q}. Here, unlike in Fig.~\ref{fig:GCB_Q}, the heat flux scaling $Q \propto q$ is identical to that derived assuming perpendicular isotropy \cite{barnes_critically_2011}.

In Figs.~\ref{fig:GCB_correlation_x} and \ref{fig:GCB_correlation_y}, we show that the correlations perpendicular to the magnetic field also obey the scalings \eqref{eq:GCB_scalings}. Here, the radial correlation function is defined as
\begin{eqnarray}
    \mathcal{C}(\Delta x) = \frac{\left\langle \delta\varphi^\mathrm{NZ}(x,y,\theta)\delta\varphi^\mathrm{NZ}(x+\Delta x,y,\theta)\right\rangle_{xy\theta}}{ \left\langle (\delta\varphi^\mathrm{NZ}(x,y,\theta))^2 \right\rangle_{xy\theta}}
\end{eqnarray}
and analogously for $\mathcal{C}(\Delta y)$. The correlation functions are dominated by the contribution from the peak in the Fourier spectrum; therefore, Figs.~\ref{fig:GCB_correlation_x} and \ref{fig:GCB_correlation_y} confirm the alignment of the outer scale in Figs.~\ref{fig:GCB_spectrum_ky} and \ref{fig:GCB_spectrum_kx}.

We show in Figs.~\ref{fig:GCB_correlation_kx} and \ref{fig:GCB_correlation_ky} that the parallel correlation lengths also follow the scalings \eqref{eq:GCB_scalings} at the outer scale. Here, the parallel correlation length as a function of $k_y$ is defined like in \cite{barnes_critically_2011}, and the parallel correlation length as a function of $k_x$ is defined analogously,
\begin{align}
    & (\overline{\Delta \theta})_{k_x} = \\\nonumber 
    & \quad \int\mathrm{d}\Delta\theta\, \frac{\sum_{k_y \neq 0} \delta\hat\varphi_{k_x,k_y}(\theta=0) \delta\hat\varphi^*_{k_x,k_y}(\theta=\Delta\theta) }{\sum_{k_y \neq 0} \abs{\delta\hat\varphi_{k_x,k_y}(\theta=0)}^2 }.
\end{align}
We note that the parallel correlation length $(\overline{\Delta \theta})_{k_x}$ has spikes at $k_x \rho_i \approx 0.5$ corresponding to the TSMs, which are correlated over approximately half a poloidal turn.

\subsection{Zonal flow amplitude scaling}

For sufficiently large $q$, there is a clear scale separation between the turbulence outer scale $k_x^o \rho_i \sim 1/q$ and the TSM scale $k_x \rho_i \sim 0.5$, where the turbulence spectrum exhibits only a small peak (see Fig.~\ref{fig:GCB_spectrum_kx}). Therefore, the scalings \eqref{eq:GCB_scalings} justify the secondary model considered in Figs.~\ref{fig:toroidal_secondary_GKsec_2D}~and~\ref{fig:TSM_physics} to describe the TSM, as the turbulence varies over long spatio-temporal scales $k_x^o \rho_i \sim 1/q$ and $\omega_\mathrm{NL}^o \sim v_{Ti}/qR$ compared to the TSM ($k_x \rho_i \sim 1$ and $\omega^\mathrm{ZF}\sim v_{Ti}/R$).

\begin{figure}[!hb]
    \centering
    \includegraphics[width=\columnwidth, trim={0.6cm 0.8cm 0.6cm 0.6cm},clip]{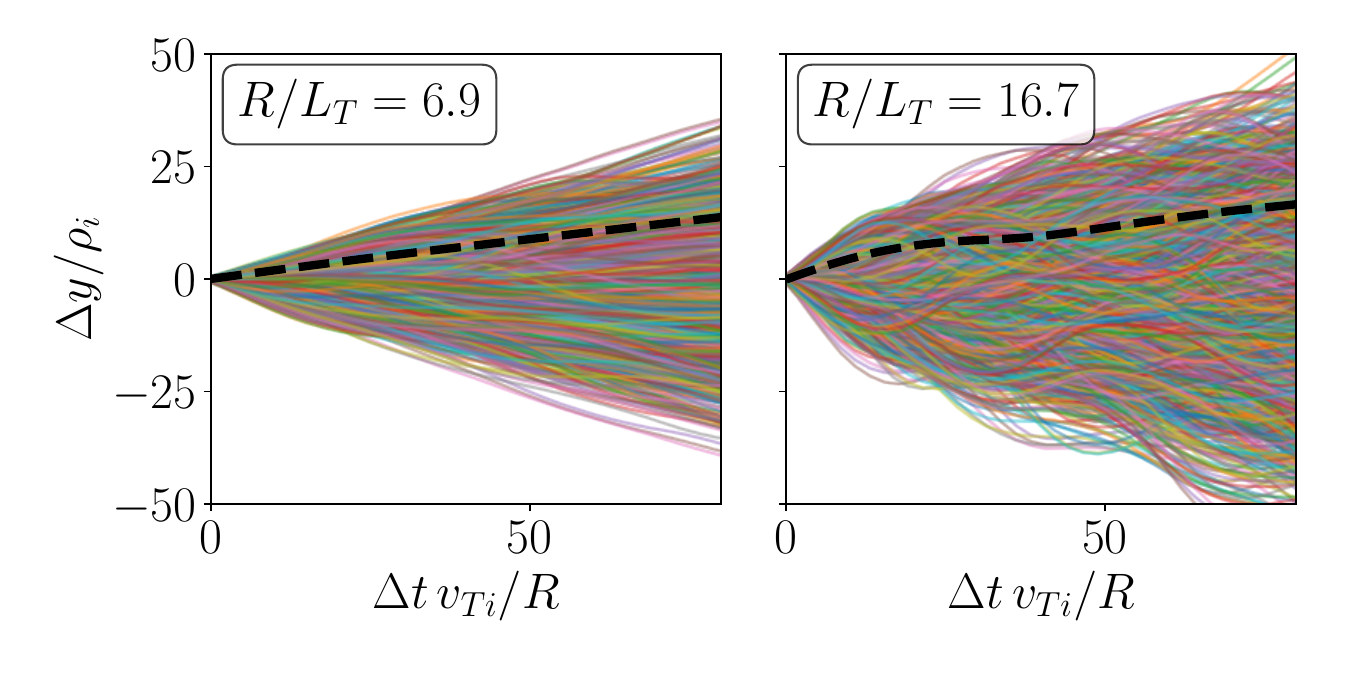}
	\caption{Displacement due to advection by the zonal flow (ZF) velocity field close to marginality (left), where stationary ZFs are dominant and coherently shear turbulent eddies, and far from marginality (right), where shearing by the TSMs is diffusive. The ZF velocity field is taken at different radial positions from fully nonlinear \texttt{stella} simulations at $q=1.4$ and different $R/L_T$ values. The black dashed line corresponds to the root-mean-square displacement.}
    \label{fig:diffusive_pic}
\end{figure}

The importance of TSMs for turbulence saturation may seem puzzling given that ZFs oscillating at a frequency $\omega^\mathrm{ZF}$ faster than the decorrelation rate $\omega_\mathrm{NL}$ have been argued to be inefficient at shearing apart turbulent eddies \cite{hahm_shearing_1999}. Intuitively, the net displacement by ZF advection cancels out over an oscillation period, curtailing the shearing effect. In fact, as shown in Fig.~\ref{fig:diffusive_pic}, the advection by the TSMs is uncorrelated in time, such that a substantial net displacement by ZF advection results from diffusion in the binormal direction $y$.

To affect turbulence saturation, the ZFs must cause diffusion over a length scale $l_y^o$ in a nonlinear time, i.e., $(l_y^o)^2 \sim D/\omega_\mathrm{NL}^o$, where $D\sim (v_E^\mathrm{Z})^2 / \omega^\mathrm{ZF}$ is the diffusivity due to ZFs. Using \eqref{eq:GCB_scalings}, the ZF energy 
\begin{equation} \label{eq:E_ZF}
    E^\mathrm{ZF} \equiv \frac{e_i^2}{2n_i T_i} \left\langle \int\mathrm{d}^3 v\; \varphi^\mathrm{Z} \left( \varphi^\mathrm{Z} - \overline{\overline{\varphi^\mathrm{Z}}} \right) F_M \right\rangle_{tx\theta}
\end{equation}
is predicted to scale as
\begin{eqnarray} \label{eq:ZF_scaling}
   \frac{E^\mathrm{ZF}}{T_i} \approx \left\langle \left(\frac{v_E^\mathrm{Z}}{v_{Ti}}\right)^2 \right\rangle_{tx\theta} \sim q \left( \frac{R}{L_T} \right)^2 \left( \frac{\rho_i}{R} \right)^2.
\end{eqnarray}
Here, $\overline{\overline{\varphi^\mathrm{Z}}}$ corresponds to two successive gyro-averages, first at fixed $\bs{R}$ and then at fixed $\bs{r}$, which in the long-wavelength limit $k_\perp \rho_i \ll 1$ give the approximation in \eqref{eq:ZF_scaling}. As shown in Fig.~\ref{fig:GCB_ZF}, the predicted ZF scaling \eqref{eq:ZF_scaling} is well satisfied by the small-scale TSMs in simulations.

\section{Discussion}
\label{sec:Discussion}

In this article, we presented the toroidal secondary mode (TSM), a new mode leading to small-scale propagating ZFs in ITG turbulence simulations (Fig.~\ref{fig:ZF_real_Fourier}). The basic physical mechanism that drives the TSM was explained to rely on the nonlinear generation of up-down pressure asymmetry (a detailed theory of TSMs may be found in \cite{nies_theory_2025-1}). The turbulence saturation level is regulated to be near the threshold amplitude $v_{Ex}\sim v_{Mx}$ above which the TSM becomes unstable (Fig.~\ref{fig:toroidal_secondary_GKsec_2D}). As a consequence, strongly-driven ITG turbulence follows the scaling laws \eqref{eq:GCB_scalings} in the background temperature gradient and the safety factor, in agreement with gyrokinetic simulation results (Figs.~\ref{fig:GCB} and \ref{fig:GCB_extra}) and experimental measurements \citep{ghim_experimental_2013}. We note that various other dimensionless parameters may affect the turbulence, such as the ratio of temperature and density gradient or the magnetic shear; such parameter dependences should be elucidated in future work.

We also note that TSMs resulting from electron-gyroradius-scale turbulence could differ considerably in character from those presented in this article, as the ZF inertia will play a more important role. For ion scale modes, the ZF inertia does not contribute significantly to the TSM mechanism because the modified adiabatic electron response in \eqref{eq:quasineutrality} gives a small ZF inertia at long wavelengths in the vorticity equation \eqref{eq:vorticity}.

The TSMs shown in Fig.~\ref{fig:ZF_real_Fourier} are reminiscent of the avalanches observed in gyrokinetic simulations \cite[e.g.][]{garbet_profile_2004, mcmillan_avalanchelike_2009}, whose physical mechanism remains debated. From Fig.~\ref{fig:ZF_real_Fourier}, the TSM is seen to have a typical propagation velocity $\sim 2 v_{Ti} \rho_i/R$, comparable to previously reported avalanche propagation speeds \cite{mcmillan_avalanchelike_2009, rath_comparison_2016}. Furthermore, the turbulence amplitude threshold required to destabilise the TSM could explain the sandpile-like behaviour associated with avalanches.

\section*{Acknowledgments}

This work was supported by U.S. DOE DE-AC02-09CH11466 and DE-FG02-93ER54197, by Scientific Discovery Through Advanced Computing (SciDAC) Grant No. UTA18000275, and by the Engineering and Physical Sciences Research Council (EPSRC) [EP/R034737/1]. The simulations presented in this article were performed on computational resources managed and supported by Princeton Research Computing, a consortium of groups including the Princeton Institute for Computational Science and Engineering (PICSciE) and the Office of Information Technology's High Performance Computing Center and Visualization Laboratory at Princeton University.

\section*{Data availability}

The data that support the findings of this article are openly available \cite{nies_dataset_2026}.

\appendix

\section{DETAILS OF GYROKINETIC SIMULATIONS}
\label{app:details_GK_sims}

All simulations presented in this article were performed on a flux tube extending for a single poloidal turn. In addition to the parameters given in the main text, a magnetic shear of $\hat s = \mathrm{d}\ln q/\mathrm{d}\ln r = 0.8$ was used, and the density gradient was held fixed at $R/L_n \equiv -R \;\mathrm{d}\ln n_i/\mathrm{d}r = 2.2$. A hydrogen plasma with $Z_i=1$ was considered, and the electron to ion temperature ratio was set to $\tau=1$.

The numerical parameters employed in the \texttt{stella} \citep{barnes_stella_2019} gyrokinetic simulations are provided in Table~\ref{tab:stella}. The quantities $L_x/\rho_i$ and $L_y/\rho_i$ denote the box size in the radial and binormal directions, respectively, and are related to the minimum wavenumber through $k_\mathrm{min} = 2\pi/L$. Furthermore, $N_\theta$ gives the parallel resolution while $N_x$ and $N_y$ denote the number of Fourier modes in the radial and binormal directions. The velocity resolution is given by $N_\mu$ and $N_{v_\parallel}$ for the magnetic moment $\mu$ and the parallel velocity $v_\parallel$, respectively, with $v_\parallel/v_{Ti} \in [-3.5,3.5]$ and $\mu \in [0, 4.5 \,m_i v_{Ti}^2/ \text{min}(B)]$. Finally, $D_h$ denotes the hyperdissipation parameter.

\begin{table}[!hb]
    \centering
    \begin{tabular}{c|c|c|c|c|c|c|c|c|c}
         $\kappa$ & $q$ & $L_x/\rho_i$ & $L_y/\rho_i$ & $N_\theta$ & $N_x$ & $N_y$ & $N_\mu$ & $N_{v_\parallel}$ & $D_h$\\ \hhline{=|=|=|=|=|=|=|=|=|=}
         6.9      & 1.4 & 156          & 157          & 24         & 256   & 192   & 12      & 64                & 0.05\\ 
         10.4     & 1.4 & 231          & 232          & 24         & 384   & 288   & 12      & 64                & 0.05\\ 
         13.9     & 1.4 & 312          & 314          & 24         & 512   & 384   & 12      & 64                & 0.05\\ 
         16.7     & 1.4 & 625          & 628          & 32         & 1152  & 1024  & 12      & 64                & 0.12\\ 
         20.9     & 1.4 & 625          & 628          & 32         & 1152  & 1024  & 12      & 64                & 0.15\\ \hline
         10.4     & 2.8 & 412          & 415          & 32         & 384   & 256   & 12      & 64                & 0.10\\ 
         10.4     & 4.2 & 500          & 628          & 32         & 512   & 384   & 12      & 64                & 0.10
    \end{tabular}
    \caption{Parameters of \texttt{stella} \citep{barnes_stella_2019} simulations.}
    \label{tab:stella}
\end{table}

As shown in Table~\ref{tab:GX}, the simulations using the \texttt{GX} code \citep{mandell_laguerrehermite_2018, mandellGXGPUnativeGyrokinetic2024} could be run with fixed numerical parameters as $\kappa=R/L_T$ and $q$ were varied, as the computational cost is small owing to the use of GPUs. In the $\kappa$ scan, temperature gradient values of $\kappa\in[6.9, 13.9, 20.9, 27.8, 34.7, 41.7, 48.6]$ were considered, at fixed safety factor $q=1.4$, while for the $q$ scan, $\kappa=13.9$ and $q \in [1.4, 2.8, 4.2, 5.6, 7.0]$. The velocity resolution is given here by the number of Hermite and Laguerre moments $N_h$ and $N_l$, respectively.

\begin{table}[!hb]
    \centering
    \begin{tabular}{c|c|c|c|c|c|c|c|c}
                        & $L_x/\rho_i$ & $L_y/\rho_i$ & $N_\theta$ & $N_x$ & $N_y$ & $N_l$ & $N_h$   & $D_h$ \\ \hhline{=|=|=|=|=|=|=|=|=}
         $\kappa$ scan  & 238          & 400          & 14         & 512   & 1024  & 12      & 8     & 0.1   \\\hline
         $q$ scan       & 318          & 533          & 14         & 512   & 1024  & 6       & 8     & 0.1 
    \end{tabular}
    \caption{Parameters of \texttt{GX} \citep{mandell_laguerrehermite_2018, mandellGXGPUnativeGyrokinetic2024} simulations.}
    \label{tab:GX}
\end{table}

\section{DERIVATION OF THE VORTICITY EQUATION}
\label{app:derivation_vorticity}

The flux-surface average and time derivative of the quasineutrality equation \eqref{eq:quasineutrality} is
\begin{eqnarray}
    \frac{T_i}{e_i n_i} \left\langle \int\mathrm{d}^3 v\; \overline{\partial_t h_i} \right\rangle_{y\theta} = \partial_t \langle \varphi \rangle_{y \theta}.
\end{eqnarray}
The left-hand-side can be simplified using the gyrokinetic equation \eqref{eq:gyrokinetic_eq_realspace_gs}; the contributions from binormal derivatives $\bs{\tilde{v}}_M \cdot \nabla y \partial_y h_i$ and $\overline{\bs{v}_E}\cdot \nabla F_{Mi} \propto \partial_y \overline{\varphi}$ vanish due to periodicity in $y$ and the parallel streaming contribution $v_\parallel \hat b \cdot \nabla h_i$ vanishes due to periodicity of zonal modes in $\theta$. We thus obtain
\begin{eqnarray} \label{eq:timeder_fsa_quasineutrality}
    \frac{T_i}{e_i n_i } &\Big\langle \int\mathrm{d}^3 v\; \Big[\frac{e_i}{T_i} F_{Mi} \partial_t \overline{\overline{\varphi}}  - \tilde v_{Mx} \partial_x \overline{h_i} - \overline{\overline{\bs{v}_E} \cdot\nabla  h_i} \Big]  \Big\rangle_{y \theta} \nonumber \\
    & =  \partial_t \langle \varphi \rangle_{y \theta}.
\end{eqnarray}

\begin{figure}[!hb]
    \centering
    \includegraphics[width=0.99\linewidth]{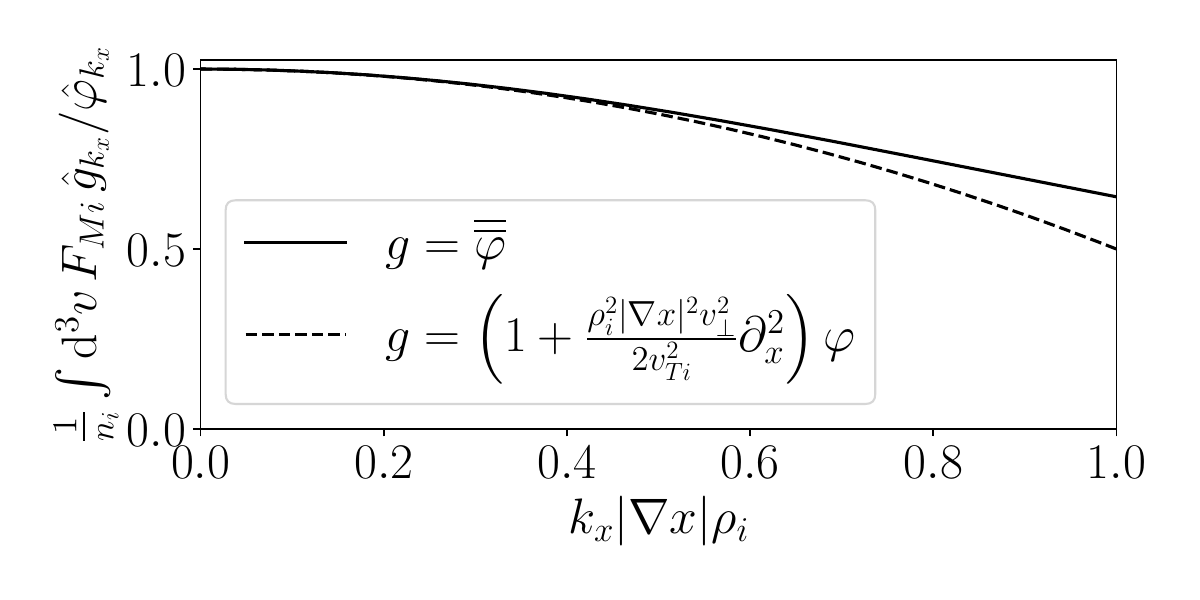}
    \caption{Expansion of gyro-averages for long perpendicular wavelengths: the velocity integral of the approximation \eqref{eq:phi_approx_LW} is shown for a Fourier mode $\varphi = \hat {\varphi}_{k_x} e^{i k_x x}$.}
    \label{fig:Gamma0}
\end{figure}

Let us us now consider long perpendicular wavelengths, where the gyro-averages in \eqref{eq:timeder_fsa_quasineutrality} may be approximated as
\begin{equation} \label{eq:phi_approx_LW}
    \langle\overline{\overline{\varphi}} \rangle_{y} \approx \langle \varphi \rangle_{y} + \left\langle \frac{\rho_i^2 \abs{\nabla x}^2  v_\perp^2}{2 v_{Ti}^2} \partial_x^2\varphi \right\rangle_{y},
\end{equation}
$\tilde v_{Mx} \partial_x \overline{h_i} \approx \tilde v_{Mx} \partial_x h_i$, and 
\begin{eqnarray} \label{eq:NL_term_LW_approx}
    &\left\langle  \overline{\overline{\bs{v}_E} \cdot\nabla  h_i} \right\rangle_y \approx \bigg\langle \bs{v}_E \cdot\nabla \overline{h_i} \\ 
    & + \frac{v_\perp^2}{4 \Omega_i^2} \left[ \nabla_\perp^2 (\bs{v}_E \cdot\nabla  h_i) + \nabla_\perp^2 \bs{v}_E \cdot \nabla h_i - \bs{v}_E \cdot \nabla \nabla_\perp^2 h_i \right] \bigg\rangle_y \nonumber.
\end{eqnarray}
These approximations typically remain accurate for perpendicular wavenumbers $k_\perp \rho_i \lesssim 0.8$, e.g. the density moment of \eqref{eq:phi_approx_LW} is plotted against $k_x$ in Fig.~\ref{fig:Gamma0}; in particular, the expansion is reasonably well justified at the typical scale of the toroidal secondary mode $k_x \rho_i \approx 0.5$.

When inserted into \eqref{eq:timeder_fsa_quasineutrality}, the first term of \eqref{eq:NL_term_LW_approx} vanishes due to quasineutrality and $\bs{v}_E \cdot \nabla \varphi = 0$, and the remaining terms may be simplified to give
\begin{eqnarray}
    & \partial_t \left\langle \rho_i^2 \abs{\nabla x}^2  \partial_x^2 \varphi \right\rangle_{y \theta} = \\
    & \frac{2 T_i}{e_i n_i} \bigg\langle \int\mathrm{d}^3 v\; \bigg[ \tilde{v}_{Mx} \partial_x h_i + \frac{v_\perp^2}{2\Omega_i^2} \nabla_\perp \cdot \left( \nabla \bs{v}_E \cdot \nabla h_i \right)  \bigg] \bigg\rangle_{y \theta} \nonumber.
\end{eqnarray}
Finally, periodicity in $y$ may again be used to simplify the last term to a radial derivative, and \eqref{eq:vorticity} then follows using the definitions $v_E^\mathrm{Z} = \abs{\nabla x} \partial_x \langle \varphi \rangle_y / B$ and $T_i = m_i v_{Ti}^2/2$.

\bibliography{references}

\end{document}